\documentclass[prl,aps,twocolumn,showpacs]{revtex4}
\usepackage{graphicx}
\begin{document}

\title{Dendritic flux avalanches and nonlocal electrodynamics in thin superconducting films}
\author{Igor S. Aranson$^{1}$, Alex Gurevich$^{2}$,
Marco S. Welling$^{3}$, Rinke J. Wijngaarden$^{3}$, Vitalii K.
Vlasko-Vlasov$^{1}$, Valerii M. Vinokur$^{1}$, and Ulrich Welp$^1
$}.
\affiliation{$^{1}$Materials Science Division, Argonne National Laboratory, Argonne, Illinois 60439\\
$^{2}$Applied Superconductivity Center, University of
Wisconsin, Madison, Wisconsin 53706 \\
$^{3}$Division of Physics and Astronomy, Faculty of Sciences, Vrije Universiteit,
De Boelelaan 1081, 1081HV Amsterdam, The Netherlands}
\date{\today}

\begin{abstract}
We present numerical and analytical studies of coupled nonlinear
Maxwell and thermal diffusion equations which describe
nonisothermal dendritic flux penetration in superconducting films.
We show that spontaneous branching of propagating flux filaments
occurs due to nonlocal magnetic flux diffusion and positive
feedback between flux motion and Joule heat generation. The
branching is triggered by a thermomagnetic edge instability which
causes stratification of the critical state. The resulting
distribution of magnetic microavalanches depends on a spatial
distribution of defects. Our results are in good agreement with
experiments performed on Nb films.
\end{abstract}
 \maketitle

Penetration of magnetic flux in a type-II superconductor can
result in nonequilibrium pattern formation, such as magnetic
macroturbulence \cite{tur}, kinetic front roughening \cite{front},
magnetic microavalanches \cite{mav}, and dendritic structures
\cite{rmp}. Dendritic flux penetration has been revealed by
magneto-optical imaging (MOI) on multiple scales $\sim 1-100 \mu$m
much greater than intervortex spacings in
YBa$_2$Cu$_3$O$_7$\cite{denYBCO}, Nb \cite{denNb1,denNb2},
Nb$_3$Sn\cite{denNb3Sn}, and MgB$_2$\cite{denMgB2}. Similarity of
these dynamic flux patterns in different materials indicates a
generic collective behavior of vortices.

Recently it has been shown both experimentally and theoretically
that dendritic flux penetration is due to a positive feedback
between moving flux and the Joule heating coupled by a highly
nonlinear voltage-current characteristic \cite{rmp,denMgB2,agv}.
The resulting thermal bistability of current-carrying
superconductors gives rise to  switching waves between a cold
superconducting phase and a hot resistive phase self-sustained by
the Joule heating \cite{gm}. Dendritic flux penetration in
superconductors has  analogs in the theory of pattern formation out
of equilibrium \cite{npat} and instability of solidification fronts
\cite{solid}.

Dendritic flux dynamics in superconductors was observed in
numerical analysis of nonisothermal magnetic diffusion in a
slab in a parallel field, when flux penetration was
triggered by a local heat pulse \cite{agv}. However, experiments
have been mostly done on films in a perpendicular ramping magnetic
field in which case magnetic flux diffusion becomes strongly
nonlocal due to long-range interaction of vortices
 \cite{ehb}. In this Letter we calculate dendritic flux penetration
controlled by nonlocal magnetic flux diffusion coupled to thermal
diffusion in thin films. We report a novel {\it nonlocal}
mechanism of flux branching, which  captures salient features of
dendritic flux penetration in superconducting films.

We consider a thin film strip of the width $w$ along the $y$-axis
and thickness $d \ll w$ in the $xy$ plane perpendicular to the
magnetic field $H_0$. Distributions of the magnetic induction,
${\bf B}({\bf r},t)$, and temperature $T({\bf r},t)$ are described
by the Maxwell equation coupled to the heat diffusion:
    \begin{eqnarray}                
    C\partial_tT=\nabla \kappa\nabla T-(T-T_0)h/d+{\bf J}  {\bf  E} (J,T),
    \label{heat} \\                 
    \partial_t {\bf B} =-\nabla\times{\bf E(J,T)},\;\;
    \nabla \times {\bf H}  ={\bf J} \delta(z).
    \label{max}
    \end{eqnarray}
Here $C(T)$ is the heat capacity, $\kappa(T)$ is the thermal
conductivity, $h(T,T_0)$ is the heat transfer coefficient to the
coolant or substrate held at the temperature $T_0$, and ${\bf E}={\bf J}E(J,T)/J$ is the
electric field, which strongly depends on $T({\bf r},t)$ and the sheet current density
${\bf J}({\bf r},t)$.

The $E(J,T,B)$ characteristic accounts for a resistive flux flow state with
$E=(J-J_c)\rho_F$ for $J>J_c$ and a low-resistive flux creep state with
$E=E_c\exp (J-J_c)/J_1$ for $J<J_c$, where $J_c(T,B)$ is the critical current density.
We use the following interpolation formula expressed in terms of observable parameters \cite{agv}:
    \begin{equation}                
    E=\rho_F J_1\ln [1+\exp (J-J_c)/J_1],
    \label{interp}
    \end{equation}
where $J_1(T)$ is logarithmic flux creep rate ($J_1\ll J_c$ below
the irreversibility field $B<B^*$), and $\rho_F(T)=\rho_n
B/B_{c2}$ is the flux flow resistivity.

We consider weak Joule heating, for which the most essential
temperature dependence comes from $E(T)$, while other parameters
may be taken at $T=T_0$. The relation between current and the
$z$-component $B_z$ in a film is given by the non-local
Biot-Savart law. Expressing $J_x=\partial_yg$ and
$J_y=-\partial_xg$ in Eq. (\ref{max}) in terms of the current
stream function $g(x,y,t)$, we obtain the equations for $g$ and
the dimensionless temperature $\theta$:
    \begin{eqnarray}
    \tau{\dot g}=\hat K \bigl[
    \partial_x[r(j,\theta) \partial_x g
    +\partial_y[r(j,\theta)\partial_y g ]  - \tau\dot H_0(t)
    \bigr],
    \label{zeta} \\             
    {\dot\theta}=\nabla^2\theta-\theta+\alpha j^2r(j,\theta).\qquad\qquad
    \label{t}            
    \end{eqnarray}
Here we define the operator $\hat K$ in the Fourier space, $\hat K
= \sum_{\bf k}\sin(k_x x)\sin(k_y y)g_{\bf k}/k$, where $k_x=\pi
n/L$ and $k_y=\pi m/w$ with integer $m$ and $n$ to ensure zero
normal component of ${\bf J}$ at edges of a rectangular film of
width $w$ and length $L$ \cite{eq}. Furthermore, $\theta =
(T-T_0)/(T^*-T_0)$, $J_c(T^*)=0$, $j=J/J_1=[j_x^2+j_y^2]^{1/2}$,
$H_z(x,y)=\sum_{\bf k} \exp (-d k-i{\bf kr})g_{\bf k}k/2+ H_0(t)$,
the factor $\exp (-d k)$ accounts for a finite film thickness, the
derivatives in Eqs. (\ref{zeta}) and (\ref{t}) are taken over
normalized time $t/t_h$ and coordinates ${\bf r}/L_h$, and the
nonlinear resistivity $r(j,\theta )=\ln [1+\exp
(j-j_c(\theta))]/j$ is obtained from Eq. (\ref{interp}). Here the
thermal length $L_h=(d\kappa/h)^{1/2}$ and time $t_h=Cd/h$ define
the spatial scale and the cool down time of $T({\bf r},t)$ at a
frozen ${\bf J}({\bf r})$. Hereafter we take
$j_c(\theta)=j_0(1-\theta)$ for $\theta < 1$, and $j_c=0$ for
$\theta>1$, and $J_1(T)=$~const \cite{agv}, assume that a uniform
magnetic field $H_0(t)={\dot H}_0 t$ is ramped up with the rate
${\dot H}_0$, and neglect the field dependence of $J_c$. Evolution
of $\theta({\bf r},t)$ and $g({\bf r},t)$ is controlled by two
dimensionless parameters:
        \begin{equation}        
        \tau=\frac{\mu_0\sqrt{d\kappa h}}{2\rho_F C},\qquad \alpha=\frac{\rho J_1^2d}{h(T^*-T_0)}.
        \label{param}
        \end{equation}
Here $\tau=t_m/t_h$ is the ratio of magnetic and thermal diffusion
times, and $\alpha$ quantifies the Joule dissipation. Magnetic
nonlocality strongly reduces $\tau =\tau_0d/2L_h$ in a film as
compared to $\tau_0=\mu_0\kappa/\rho_F C$ in the bulk. Indeed,
flux diffusion over a distance $L_h$ along a film takes $t_m\sim
dL_h/D_m$, while thermal diffusion takes $t_h\sim L_h^2/D_h$,
where $D_m=\mu_0/\rho_F$ and $D_h=\kappa/C$. For Nb films with
$d=0.5\mu$m, at $4.2$K, ($\kappa\simeq 0.2$W/cmK, $h\simeq$
1W/cm$^2$K, $\rho_n\simeq 3\cdot10^{-7}\Omega$cm \cite{denNb2}, and $C\simeq
2\cdot10^{-3}$J/cm$^3$K), we obtain $L_h=(d\kappa/h)^{1/2}\simeq
0.03$mm, $t_h=Cd/h\simeq 10^{-7}$s, $d/L_h \sim 10^{-2}$, $\tau
\sim 0.1$, with $\tau$ decreasing as $T_0$ increases.

We used Eqs. (\ref{zeta}), (\ref{t}) to study flux penetration in
a film with periodic boundary conditions (b.c.) along $x$ and
$\partial_y\theta=g=0$ at $y=0,w$. Eqs. (\ref{zeta}), (\ref{t})
were solved numerically by a quasi-spectral method based on the
Fast Fourier Transform; up to $1024\times512$ harmonics were used.
To implement non-periodic b.c. in the $y$-direction we used the
domain of doubled length with the condition $g(x,y)=-g(x,2w-y)$.
Calculated steady-state distributions of $B_z(x)$ are very close
to those of the Bean model \cite{net}. In the majority of
numerical runs for $\tau\ll 1$, we observed spontaneous avalanches
induced by ramping magnetic field $H_0(t)={\dot H}_0t$, starting
from a zero field cooled state. We also took into account
randomly-distributed macroscopic defects modelled by $j_c({\bf r}
) =j_{0}[1-\theta- \sum_{i} q_i$$ \cosh^{-1}(|{\bf r} -{\bf
r}_i|/\xi_0)]$ where $q_i$, ${\bf r}_i$ and $\xi_0$ determine the
strength, the position and the radius of the i-th defect. Extended
current-blocking defects have been commonly revealed by the MOI
\cite{mo}.

\begin{figure}[ptb]
\includegraphics[width=2.6in,angle=0]{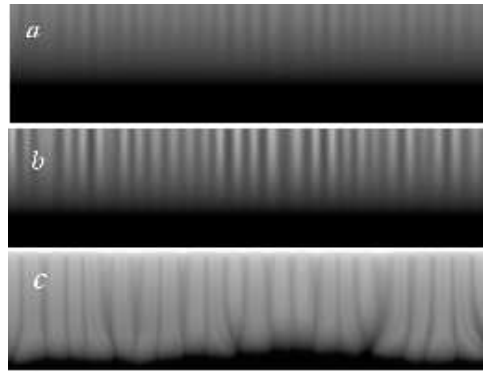}
\caption{Temperature distributions in a film with no defects for
${\dot H}_0=5J_1/t_h$, $\tau=0.0025$, $\alpha=0.008$, system size
$600L_h\times150L_h$ (only quarter part is shown) and $j_0=20$ at
$t=16.75t_h$ (a) and $t=17.75t_h (b)$; (c) magnetic flux pattern
for $t=20t_h$} \label{Fig1}
\end{figure}

We start with flux penetration in a film with no macro-defects,
Fig.~\ref{Fig1} and Movie 1 in \cite{net}. The ramping magnetic
field first caused penetration of a stable cold flux front. Then
an instability causing {\it periodic modulations} of temperature
and propagation of hot magnetic filaments over the preceding
smooth magnetic flux distribution develops at the film edge. Once
the first wave of magnetic filaments reaches the central line
where the magnetization currents change direction, the filaments
widen and start splitting at the ends, similar to that of a slab
in a parallel field \cite{agv}. At the same time, a second wave of
hot filaments start propagating from the film edge mostly between
the paths of the filaments of the first wave. The new filaments
are wider and exhibit shape distortions due to interaction with
preceding filaments. Eventually the film cools down and the flux
penetration stops, resulting in a frozen multi-filamentary
structure \cite{net}.

To address the mechanism of the edge instability, we performed a
linear stability analysis of steady-state distributions $T(y)$ and
$E_0(y)$. Because $E(J,T)$ depends on $J$ and $T$,
linearization of Eqs. (\ref{zeta}) and (\ref{t}) gives two coupled
equations for the perturbations $\delta\theta$ and $\delta g$, which
depend on $E_0(y)$ and $T(y)$. The problem can
be simplified by the fact that the flux stratification period
$\ell$ in Fig. 1 is much smaller than the film width $w$. Because
both $E_0(y)$ and $T(y)$ vary on the scale $\sim w$, we can
neglect slow variations of $E_0(y)$ and $T(y)$,
taking uniform $T=T_0$ and $E_0=\dot{B}_0w/2$ at the film
edge. Then the solutions, which satisfy the boundary conditions
$\delta J_y=\partial_y\delta T=0$ at the film edge ($y=0$), take
the form $\delta\theta\propto e^{\lambda t+ikx}\cos qy$, and
$\delta g\propto e^{\lambda t+ikx}\sin qy$. From Eqs.
(\ref{zeta}), (\ref{t}), we obtain the following dispersion
relation:
    \begin{eqnarray}
        (\lambda+1+q^2+k^2-\beta)(\lambda\tau_i\sqrt{k^2+q^2} +q^2+sk^2) \nonumber \\
        +q^2\beta (1+s)=0, \qquad\qquad\quad
    \label{disp}        
    \end{eqnarray}
where $\lambda$ and $(k,q)$ are measured in $t_h^{-1}$, and
$L_h^{-1}$, respectively. The dissipation control parameter
$\beta=(dJ/h)\partial E/\partial T$ depends on $E_0$: if
$E(J)=E_c\exp[(J-J_c)/J_1]$ then $\beta\simeq E_0J_cd|\partial
J_c/\partial T|/J_1h$ for $J<J_c$. The parameter
$\tau_i=\mu_0\sqrt{dh\kappa}/2\rho C$, is similar to that in Eq.
(\ref{param}), except that $\rho_F$ is now replaced with the
differential resistivity $\rho (E)=\partial E/\partial J$, and
$s=E/J\rho (E)\simeq J_1/J_c$ is the flux creep rate.

Eq. (\ref{disp}) describes coupled thermal and magnetic diffusion
modes, the factor $\sqrt{k^2+q^2}$ accounting for the magnetic
nonlocality. Positive eigenvalues $\lambda(k,q)$ correspond to
unstable modes resulting in spontaneous thermomagnetic structures \cite{rakh}.
For slow flux diffusion $\tau\gg 1$, thermal perturbations with
$\lambda\simeq \beta-1-k^2-q^2$ are unstable above the thermal
runaway threshold, $\beta >1$. In this case $\lambda$ is maximum
at $k=q=0$ so no periodic structures are expected. A completely
different situation occurs for fast flux diffusion $\tau\ll 1$ for
which Eq. (\ref{disp}) yields
    \begin{equation}        
    \lambda=\beta-1-q^2-k^2-\frac{q^2\beta (1+s)}{q^2+sk^2}.
    \label{lamm}
    \end{equation}
The spectrum of $q$ is determined by the full set of boundary
conditions in the theory of flux jumps \cite{fj}. For further qualitative analysis
we take $q\simeq \pi  /2b$, where $\delta J_x(x,b)=0$,
$b=[1-1/\cosh(B_0/B_p)]w/2$ is the width of the flux penetrated
critical state region, and $B_p=\mu_0J_c/\pi$ \cite{perp}. For a
given $q$, the increment $\lambda(q,k)$ passes through a maximum
at the wave vector $k_m$, which defines the period $\ell = 2\pi/k_m$ of the fastest growing
thermomagnetic structure along the film. Here $sk_m^2=[q^2\beta s(1+s)]^{1/2}-q^2$, thus
$\lambda(k,q)$ is maximum at the finite $k_m$ if $\beta >\beta_i=q^2/s(1+s)$, or ${\dot B}_0>{\dot B}_i$.
For $s\ll 1$, we obtain ${\dot B}_i$ and $\ell$ in normal units:
    \begin{eqnarray}
    {\dot B}_i=\frac{\pi^2\kappa}{4b^3|\partial J_c/\partial
    T|},\qquad
    \ell^2\simeq \frac{16b^2s}{({\dot B}_0/{\dot B}_i)^{1/2}-1}.
    \label{ell}
    \end{eqnarray}
The period $\ell$ decreases as $1/s$ and ${\dot B}_0$ increase,
in agreement with our numerical results \cite{net}. The branching instability with $k_m>0$ and Re $\lambda(\beta_c,k_m,q)>0$
occurs at $\beta=\beta_c$ in Eq. (\ref{lamm}), that is, ${\dot B}_0>{\dot B}_c=hJ_1\beta_c/dbJ_c|\partial J_c/\partial T|$, where
    \begin{equation}        
    \beta_c^{1/2}=(1+2q^2)^{1/2}+q\sqrt{1+1/s}.
    \label{betac}
    \end{equation}
For $\tau_i\ll 1$, Eq. (\ref{betac}) gives $\beta_c>\beta_i$ for all $q(t)=\pi/2b(t)$, so
a thermomagnetic structure with the finite period $\ell(\beta_c)\ll b$ develops as
the width $b(t)$ exceeds $b_c$,  where ${\dot B}_0={\dot B}_c(b_c)$.
Notice that Eq. (\ref{disp}) defines a region $\tau_1<\tau<\tau_2$
in which $\lambda$ is complex, which manifests itself in
temporal oscillations of growing flux structures.

Next we consider dendritic flux penetration initiated by
macroscopic defects, both at the film edge and in the bulk. Such
defects can trigger local flux jumps even if the critical state in
the bulk is stable {\cite{fjf}} and cause branching instability of
flux filaments in a slab in a parallel field \cite{agv}. Selected
results of flux patterns in a film are shown in Fig. 2. For
$j_0=1/s=20$ in Fig. 2a, edge defects produce flux fingers
superimposed with a smooth flux front. This behavior is
characteristic of any superconductor with a highly nonlinear
$E(J)$, for which a defect of size $\xi_0$ produces a much larger
disturbance $\simeq \xi_0/s$ across the current flow \cite{ag}.
The flux fingers widen and split at the ends as they collide with
the central line where magnetization currents change direction
\cite{agv}. As $j_0$ increases, hot flux filaments in Figs. 2b and
2c get thinner and start branching even {\it before} they reach
the central line. Then new fingers start growing between the
defects due to the edge instability considered above, see
animations for more details \cite{net}. Moreover, in wider samples
(Fig. 2 c, d) the finger undergoes multiple branching giving rise
to the characteristic flux dendrites. This new branching
mechanism, which was not observed in simulations of flux patterns
in a slab \cite{agv}, is principally due to nonlocal flux
diffusion in films. Thinning the filaments as $s$ decreases
follows from Eq. (\ref{ell}), while the branching shape
instability is facilitated by magnetic nonlocality, which results
in a weaker damping of short-wave electromagnetic modes $\lambda_m
\propto k$ as compared to local flux diffusion for which
$\lambda_m\propto k^2$. Another new effect at higher values of
$j_0$ is a ``giant'' flux avalanche in Fig. 2c, which starts
propagating from the region with no surface defects after the
first wave of smaller flux filaments reached the center.

Fig. 2d shows flux patterns in a film with randomly-distributed
bulk defects. In addition to the branching due to magnetic
nonlocality, propagating flux filaments can undergo splitting
caused by local transient heat spikes as they collide with
defects. This results in local shape instability of the filaments
and their subsequent branching similar to that obtained for local
magnetic flux diffusion \cite{agv}. With the decrease of $j_0$
(which is equivalent to an increase of $T_0$ or $H_0$ in the
experiment) flux filaments become wider and eventually start
overlapping, forming a continuous flux front. However, even in
this case a significant front roughness still persists both due to
a size distribution of individual filaments and due to local heat
releases as the flux front collides with defects \cite{net}.

\begin{figure}[ptb]
\includegraphics[width=2.6in,angle=-90]{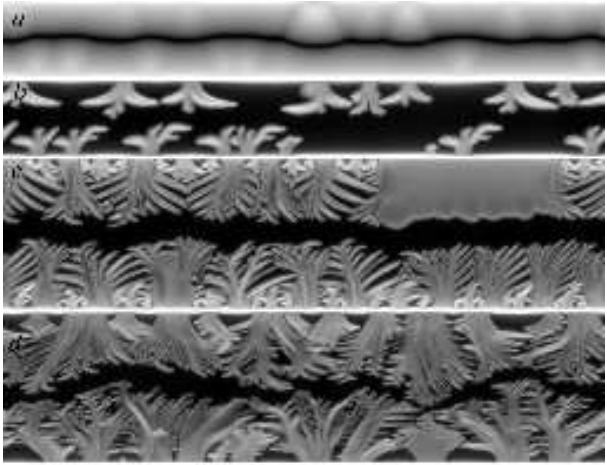}
\caption{Flux penetration in a film of $320L_h\times40L_h$ with 30
random edge defects with $\xi_0=0.25L_h$, and amplitudes $q_i$
uniformly distributed between $0$ and $0.4$, for $\dot H_0= 3
J_1/t_h$, $\alpha=0.08,\tau=0.0025$ and $t=25 t_h$ for $j_0=20 $
(a) and $j_0=80$. Black and white correspond to the Meissner and
vortex phases, respectively; (c) Flux penetration in wide film of
$600L_h\times150L_h$ for $\dot H_0= 5 J_1/t_h$, $j_0=60$, 20 edge
defects. The ``giant'' avalanche develops in a defect free region;
(d) Flux pattern in a film with 500 randomly distributed bulk
defects at $j_0=80$. See also Movies 2-6 in \cite{net}.}
\label{Fig2}
\end{figure}

\begin{figure}[ptb]
\includegraphics[width=1.7in,angle=-90]{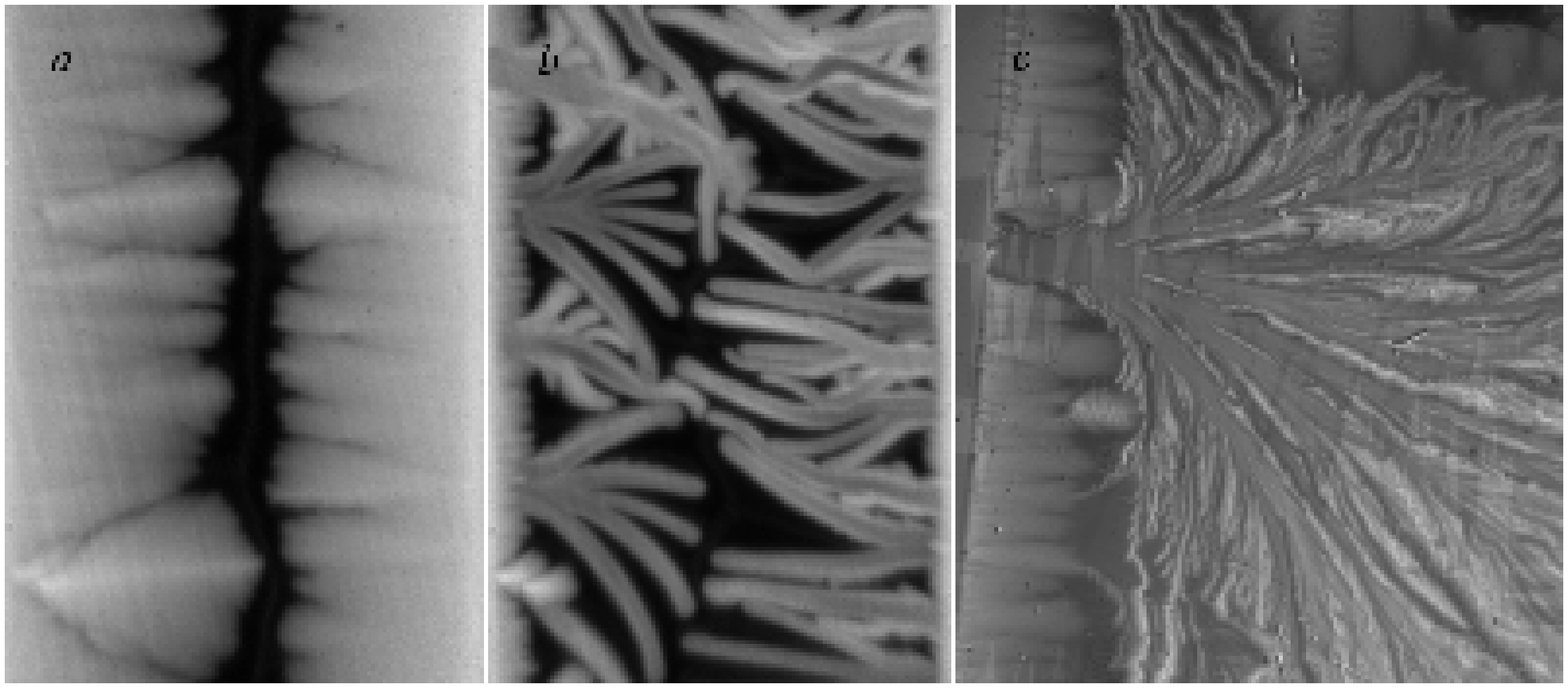}
\caption{MOI of flux branching in Nb films of Ref. \cite{denNb2}
at T=6.2K and $B_0$=31.2mT (a) and 4.7K and $B_0$=36.5mT
(b).``Giant flux avalanche'' at 4.5K in Nb film of \cite{Vit}
(c). See also Movies 7-8 in \cite{net}. } \label{Fig3}
\end{figure}

The dendritic flux penetration can be regarded as
{\it  avalanches} of vortex bundles which do not trigger a global
flux jump in the whole sample. Such avalanches produce local
temperature spikes, partial flux penetration
and a step on magnetization curves $M(B)$ \cite{mav,rmp}. Nonlocal
electrodynamics of films strongly facilitates branching flux propagation,
which requires faster magnetic diffusion
$\tau=\tau_0\sqrt{dh/\kappa}\ll 1$ characteristic of thin films with
$dh\ll \kappa$. Notice that because $b\sim
w(B_0/B_p)^2$ for small $B_0$, both critical ramp rates ${\dot
B}_i\propto B_0^{-6}$ and ${\dot B}_c\propto B_0^{-6}$ strongly
depend on the applied field $B_0$ for $B_0\ll B_p$ but level off
for $B_0 >B_p$ as $b\to w/2$. Furthermore, both
 ${\dot B}_i$ and ${\dot B}_c$ increase as $T_0$ increases.
Thus,  for a given ${\dot B}_0$, the branching  occurs at lower T above a
certain field $B_i(T)$, in agreement with many
experiments \cite{denYBCO,denNb1,denNb2,denNb3Sn,denMgB2}.

Fig. 3 shows MOI of flux penetration in two different Nb films in
ramping fields. One $9$mm$\times1.8$mm$\times0.5\mu$m Nb film
described in detail elsewhere \cite{denNb2} exhibits flux patterns
similar to those in Figs. 2a,b where dendritic flux penetration is
initiated by surface defects. For this film ($\kappa=0.2$W/cmK and
$h=1$W/cm$^2$K), we obtain $dh/\kappa\sim10^{-4}$ and the thermal
length $L_h=(d\kappa/h)^{1/2}=0.03$mm, much smaller than the film
width $w=1.8$mm (i.e. the film width is about $60L_h$ as in Figs.
2a,b), in which case the magnetic nonlocality does play the key
role. Fig. 3c shows MOI of an effectively wider
($4$mm$\times4$mm$\times0.1\mu$m) film \cite{Vit} ($w\sim 10^2L_h$),
in which in addition to small microavalanches near the film
edge a ``giant'' avalanche, similar to that triggered by a laser
pulse in YBa$_2$Cu$_3$O$_7$ \cite{denYBCO} and those in Fig. 2c,d
develops.

In conclusion, we proposed a mechanism of  flux fragmentation in
superconducting films caused by coupling of nonlocal
flux  diffusion with local thermal diffusion.
This work was supported by the NSF MRSEC (DMR 9214707) (AG); US
DOE, BES-Materials Sciences (\# W-31-109-ENG-38) (IA,VV, VKV,UW);
and  by Stichting voor Fundamenteel Onderzoek der Materie, which
is financially supported by Nederlandse Organisatie voor
Wetenschappelijk Onderzoek (MW,RW).

\end{document}